\documentclass[%
 reprint,
 amsmath,amssymb,
 aps,
]{revtex4-1}

\usepackage{graphicx}% Include figure files
\usepackage{dcolumn}% Align table columns on decimal point
\usepackage{bm}% bold math
\begin{document}

\preprint{APS/123-QED}

\title{Isospin observables from fragment energy spectra}

\author{T. X. Liu}
\altaffiliation[Present address: ]{Medical Center, University of Mississippi, Mississippi 39213}
\author{W. G. Lynch}
\email[Corresponding author: ]{lynch@nscl.msu.edu}
\author{R. H. Showalter}
\author{M. B. Tsang}
\author{X. D. Liu}
\altaffiliation[Present address: ]{SMPS Inc, 417 4th street, Marysville, California 95901}
\author{W. P. Tan}
\altaffiliation[Present address: ]{ Department of Physics, University of Notre Dame, Indiana 46556}
\author{M. J. van Goethem}
\altaffiliation[Present address: ]{Paul Scherrer Institut, Villigen, Switzerland}
\author{G. Verde}
\altaffiliation[Present address: ]{Helmholtz-Zentrum Dresden-Rossendorf, 01314 Dresden, Germany}
\author{A. Wagner}
\altaffiliation[Present address: ]{INFN, Sezione di Catania, I-95123 Catania, Italy}
\author{ H. F. Xi}
\author{ H. S. Xu}
\altaffiliation[Present address: ]{Institute of Modern Physics, CAS, Lanzhou 730000, People’s Republic of China}
\affiliation{%
National Superconducting Cyclotron Laboratory and Department of Physics and Astronomy, Michigan State University, East Lansing, Michigan 48824, USA
}%

\author{M. A. Famiano}
\affiliation{%
Department of Physics, Western Michigan University, Kalamazoo, Michigan 49008, USA
}%
\author{R. T. de Souza}
\author{V. E. Viola}
\affiliation{%
Department of Chemistry and IUCF, Indiana University, Bloomington, Indiana 47405, USA
}%
\author{R. J. Charity}
\author{L. G. Sobotka}
\affiliation{%
Department of Chemistry, Washington University, St. Louis, Missouri 63130, USA
}%

\date{\today}% It is always \today, today,
             %  but any date may be explicitly specified

\begin{abstract}
The energy spectra of light charged particles and intermediate mass fragments from $^{112}$Sn+$^{112}$Sn and $^{124}$Sn+$^{124}$Sn collisions at an incident energy of E/A=50 MeV have been measured with a large array of Silicon strip detectors. We used charged particle multiplicities detected in an array with nearly 4$\pi$ coverage to select data from the central collision events. We study isospin observables analogous to ratios of neutron and proton spectra, including double ratios and yield ratios of t/$^3$He and of asymmetries constructed from fragments with Z=3-8. Using the energy spectra, we can construct these observables as functions of kinetic energy. Most of the fragment asymmetry observables have a large sensitivity to sequential decays.
%\begin{description}
%\item[Usage]
%Secondary publications and information retrieval purposes.
%\item[PACS numbers]
%May be entered using the \verb+\pacs{#1}+ command.
%\item[Structure]
%You may use the \texttt{description} environment to structure your abstract;
%use the optional argument of the \verb+\item+ command to give the category of each item. 
%\end{description}
\end{abstract}

\pacs{Valid PACS appear here}% PACS, the Physics and Astronomy
                             % Classification Scheme.
%\keywords{Suggested keywords}%Use showkeys class option if keyword
                              %display desired
\maketitle

%\tableofcontents

\section{\label{sec:intro}Introduction\protect}

The nuclear Equation of State (EOS) describes the relation between pressure, density, temperature, and isospin asymmetry for infinite nuclear matter. The EOS has direct implications for fundamental properties of nuclei such as nuclear masses as well as the dynamics in nuclear collisions. It also affects properties of exotic astrophysical objects such as the evolution of supernovae and neutron stars. For aspects of objects with a large neutron excess, such as the crust of neutron stars and the neutron skin of Pb nuclei, understanding the EOS of asymmetric matter is very important.

Theoretical studies have shown that the EOS of asymmetric nuclear matter can be approximately expressed as \cite{1Li, 2Li}:
\begin{equation}
E(\rho, \delta) = E(\rho, \delta = 0) + S(\rho) \delta^2
\label{eq:EOS}
\end{equation}
where $\rho$=$\rho_n+\rho_p$ is the baryon density, $\delta$=($\rho_n-\rho_p$)/($\rho_n+\rho_p$) is the relative neutron excess or asymmetry of the system, and E($\rho$, $\delta=0$) is the energy per particle in symmetric nuclear matter. Significant constraints have already been placed on  E($\rho$, $\delta=0$) at high densities \cite{3Dan}. The bulk symmetry energy is denoted by S($\rho$). Its value at normal density, $S_0 \equiv S(\rho_0)$, is known to be in the range of 27-36 MeV \cite{4Dutra}.

Different density dependences of S($\rho$), which describes the sensitivity of the EOS to the difference between neutron and proton densities, depend on the nuclear forces used in the calculations \cite{4Dutra}. The experimental constraint on the isospin asymmetry term of the EOS has been poorly determined until recently \cite{5Tsang, 6Tsang}. Better knowledge of this term is essential to understand the binding energy \cite{7Myers} and the difference between neutron and proton radii in neutron-rich nuclei \cite{8Brown, 9Piek, Tsangarxiv} as well as the internal structure of neutron stars \cite{10Pethick, 11Lattimer}.

Since the isospin-dependent mean fields are opposite in sign for neutrons and protons, the reaction dynamics of neutrons and protons are affected differently, which leads to possible differences in their yields and energy spectra. The most natural observable to extract information about the asymmetry term of the nuclear EOS in Heavy Ion Collisions \cite{2Li} is the measurement of emitted neutron and proton spectra. Given the difficulties in extracting good quality neutron spectra \cite{12Famiano}, alternative observables have been suggested including the ratios of mirror charged isotopes such as t/$^3$He \cite{13Lu, 14Geraci} or $^7$Li/$^7$Be pairs and general N/Z \cite{15Liu, 16Das, 17Baran} ratios of the emitted fragment isotopes.

This paper focuses on complementary information to the neutron to proton ratios, using isospin observables constructed from the energy spectra of fragments of the  Z=1 to 8 isotopes produced in Sn+Sn collisions. We report a complete overview of measurements of fragment kinetic energies and yield ratios with particular attention to how the N/Z degree of freedom can play a key role in the dynamics of the studied reaction systems. Due to conflicting interpretations of the data obtained from different transport codes, the main goal of this paper is not to compare the data to calculations. Rather, our goal is to describe how the data are obtained, what observables have been measured in the central collisions of Sn isotopes at incident energy of 50 MeV per nucleon, and whether these observables are affected by secondary decays.

We describe the experimental set up in the next Section. Determination of the impact parameters of the Sn+Sn collisions using charged particle multiplicities is described in detail in Section \ref{sec:impactparam}. For the analysis shown in subsequent sections, a reduced impact parameter cutoff of $b/b_{max} < 0.2$ is chosen for central $^{112}$Sn+$^{112}$Sn and $^{124}$Sn+$^{124}$Sn collisions. In Section \ref{sec:fragspectra} the fragment energy spectra and mean measured energies of isotopes of all elements up to Z=8 are examined. Section \ref{sec:t3he} compares t/$^3$He yield ratios for the two symmetric Sn collision systems and contrasts the data with n/p ratios from a previous experiment \cite{12Famiano}. The average asymmetry $<$N/Z$>$ and $\Sigma$N/$\Sigma$Z constructed from fragments (from Z=3 to Z=8) of each system are examined in Section \ref{sec:fragasym}; here we note a difference in the $\Sigma$N/$\Sigma$Z distributions of the two Sn systems when isotopes of Be are either included or omitted. The findings are summarized in Section \ref{sec:summary}.

\section{\label{sec:setup}Experimental Setup}

The experiment was performed at the National Superconducting Cyclotron Laboratory at Michigan State University with beams from the K1200 cyclotron. $^{112}$Sn+$^{112}$Sn, $^{124}$Sn+$^{112}$Sn, $^{112}$Sn+$^{124}$Sn, and $^{124}$Sn+$^{124}$Sn collisions were measured by using 50 MeV per nucleon $^{112}$Sn and $^{124}$Sn beams impinging on 5 mg/cm$^2$ $^{112}$Sn and $^{124}$Sn targets. For central collisions, the results from the two mixed systems, $^{124}$Sn+$^{112}$Sn and $^{112}$Sn+$^{124}$Sn, are similar and the properties of the yields and energy spectra can be interpolated from the heavy $^{124}$Sn+$^{124}$Sn and light $^{112}$Sn+$^{112}$Sn reactions. Unless specified otherwise, only results from the measurements of the symmetric Sn+Sn systems are discussed in this paper.

Isotopically resolved particles with $3 \leq Z \leq 8$ were measured with the Large Area Silicon Strip Detector Array (LASSA) \cite{18Davin,19Wag}, an array consisting of nine telescopes, each comprised of one 65 $\mu$m thick and one 500 $\mu$m thick Si strip detector, followed by four 60 mm thick CsI(Tl) detectors. The 50$\times$50 mm$^2$ area of each LASSA telescope is divided by the strips of the second silicon detector into 256 (3$\times$3 mm$^2$) square pixels, providing an angular resolution of about $\pm$0.43$^{\circ}$. The LASSA was centered at a polar angle of $\theta_{lab}=32^{\circ}$ with respect to the beam axis, providing coverage at polar angles of $7^{\circ}\leq \theta_{lab} \leq 58^{\circ}$. At other angles, charged particles were detected in 188 plastic scintillator - CsI(Tl) detectors of the Michigan State University Miniball and Washington University Miniwall array \cite{20Souza, 21Sta}, which together subtended a range of polar angles of $7^{\circ}\leq \theta_{lab} \leq 160^{\circ}$. The Miniball/Miniwall array provided isotopic resolution for H and He nuclei and elemental resolution for intermediate mass fragments (IMF’s) with $3 \leq Z \leq 20$. The total charged particle multiplicity detected in the two arrays was used for impact parameter determination.

\section{\label{sec:impactparam}Impact Parameter Determination}

The impact parameter, $b$, of a nucleus-nucleus collision is the distance between the classical straight-line trajectories describing the initial velocities of the two nuclei before their collision. The outcome of a collision depends strongly on the impact parameter; the two nuclei will be violently disrupted for $b<<R_{proj}+R_{targ}$ but essentially undisturbed for $b>>R_{proj}+R_{targ}$. This phenomenon is explained by a simple picture wherein the charged particle multiplicity depends on the energy transferred from the relative motion of the nuclei to the internal degrees of freedom in the region where the projectile and target overlap. More charged particles will be emitted in central collisions with a stronger overlap than in peripheral collisions.

Based on the assumption that charged particles decrease monotonously with impact parameter multiplicity, a reduced impact parameter, or $\hat{b}=b/b_{max}$, can be calculated from the charged particle multiplicity. The reduced impact parameter, first suggested in ref. \cite{22Cavata}, can be written as:
\begin{equation}
\hat{b}(N_C) =  \frac{b(N_C)}{b_{max}} = \bigg[\sum^{\infty}_{N_C} P(N_C)\bigg]^{1/2} \big/ \bigg[\sum^{\infty}_{N_C(b_{max})} P(N_C)\bigg]^{1/2}
\label{eq:bhat}
\end{equation}
where $N_C$ is the charged particle multiplicity corresponding to $b(N_C)$. Here $P(N_C)$ is the relative frequency of events detected with the charged particle multiplicity equal to $N_C$, and $b_{max}$ is the average impact parameter corresponding to data taken with minimum bias multiplicity $N_C(b_{max})$ \cite{23Phair, 24Kim}. The accuracy of this relationship depends on two assumptions: First, the nucleus-nucleus cross section can be well approximated by the geometrical cross section: $\sigma_g=\pi b^2$. This is more precise at relativistic bombarding energies where the multiplicities are larger \cite{22Cavata} than at a lower bombarding energy of E/A=50 MeV. Eq. \ref{eq:bhat} is also more precise when one assumes a monotonic correlation between multiplicity and impact parameter without dispersion. At lower energies of E/A$\approx$50 MeV, however, non-negligible fluctuations in the charged particle multiplicity are expected even for collisions of well-defined impact parameter. Keeping these concerns in mind, we use Eq. \ref{eq:bhat} and group the events into distinct different bins corresponding to ``central'', ``mid-central'' and ``peripheral'' collisions \cite{25Liu}. In the following discussions, a central collision bin corresponding to $\hat{b}<0.2$ is applied to the data.

In our analysis, the charged particle multiplicity, $N_C$, consists of all the charged particles detected in the Miniball/Miniwall array and LASSA telescopes. Those detected charged particles not only include the identified particles but also unidentified particles. For the Miniball/Miniwall array, such unidentified particles include heavy fragments that stop in the fast plastic and light particles that punch through the CsI(Tl) crystals. For LASSA telescopes, the unidentified particles include heavy particles that stop in the silicon detectors, light particles that punch through the CsI(Tl), and particles that hit in the gaps between the CsI(Tl) detectors.

\begin{figure}
\includegraphics[width=0.5\textwidth]{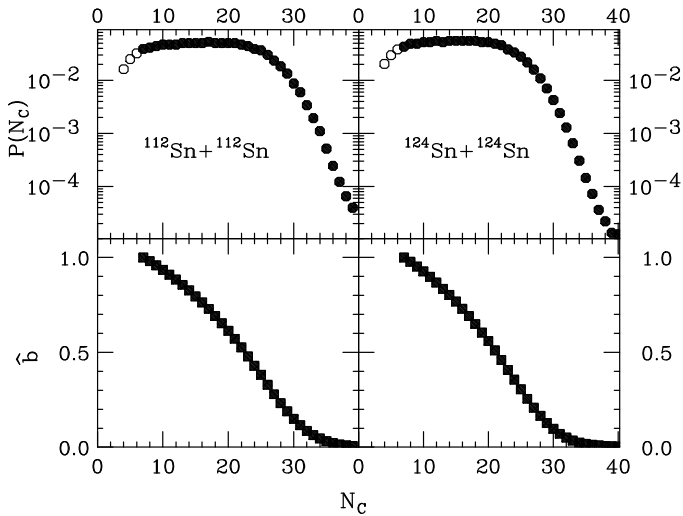}%fig1 goes here.
\caption{\label{fig:bdist}Top panels: Probability distribution of the charged particle multiplicity. Bottom panels: the corresponding reduced impact parameters as a function of the charged particle multiplicity. Data with $N_C<7$ (open circles  in the upper panel) suffer efficiency problems and are excluded from analysis.}
\end{figure}

In Fig. \ref{fig:bdist}, the probability distributions of the charged particle multiplicity as well as the reduced impact parameters are shown for $^{112}$Sn+$^{112}$Sn (left panels) and $^{124}$Sn+$^{124}$Sn (right panels) reactions. The probability distribution of $N_C$ (top panels) are rather flat between $N_C=7$ and $N_C=22$. For higher $N_C$, the probability decreases exponentially. Even though the minimum bias data was taken at $N_C=4$, we notice that data with $N_C<7$ (open points in the upper panels) suffer efficiency problems as evidenced by the sharp drop of the probability in the distributions of low $N_C$. Unlike previous studies published on this experiment \cite{26Liu}, we choose $N_C=7$ as the minimum trigger. Since $b(N_C)$ is calculated using a partial sum from $N_C$, it can be calculated independently from where the minimum $N_C$ lies.

The bottom panels of Fig. \ref{fig:bdist} show the reduced impact parameter (closed squares) calculated from the procedure described in Eq. \ref{eq:bhat}. From $\hat{b}$ we obtained the impact parameter $b= \hat{b} \times b_{max}$ using values for $b_{max}$ obtained from the measured geometric cross section $\sigma_g=\pi b^2$ for events with $N_C \geq 7$. The cross section for $N_C \geq 7$ was calculated using the relationship
\begin{equation}
\sigma = N_{event}/(N_{beam} \times N_{target})
\label{eq:sigma}
\end{equation}
where $N_{event}$ is the number of events with $N_C \geq 7$, $N_{beam}$ is the number of beam particles, and $N_{target}$ is the number of target nuclei per unit surface area. Here, $N_{target}$ was determined to 5\% accuracy from measuring the mass and area of the target. The number of events was measured precisely by the Miniball/Miniwall array and LASSA telescopes with statistical uncertainty. The number of beam particles, $N_{beam}$, was determined from the measured beam current in the Faraday cup at the end of the beam line. Due to the extremely small beam intensities used in this measurement, there may be a systematic uncertainty in the determination of the beam current. Direct counting of the beam particles with a plastic scintillator would be preferable and would provide the desired precision of a few percent at low beam intensities.

We obtained values of $b_{max}$=7.5$\pm$1.6 fm for the $^{112}$Sn+$^{112}$Sn reaction and $b_{max}$=7.0$\pm$1.4 fm for the $^{124}$Sn+$^{124}$Sn reaction. Given the experimental uncertainties in determining the absolute impact parameter, we assume $b_{max}$ =7.25 fm for both reactions. In contrast to the uncertainties associated with the absolute values for $b_{max}$, information based on the reduced impact parameter, $\hat{b}$, does not suffer from such uncertainties, but it does depend on the acceptance of the detection apparatus.

\section{\label{sec:fragspectra}Fragment Energy Spectra}

In this section, we present isotopically resolved energy spectra for fragments emitted at center of mass angles of $70^{\circ} \leq \theta_{c.m.} \leq 110^{\circ}$ in central collisions. At these angles, the coverage of the LASSA array is excellent. However, there was no detector coverage at small laboratory angles of $\theta_{lab}<7^{\circ}$ which correspond to low energy particles of less than 0.2 MeV per nucleon in the center of mass frame. We estimated the small contributions of these low energy particles. We also remove background counts that arise from particles passing out of the active regions of the telescope without being properly identified and corrected for coincidence summing of two or more particles detected within a single CsI(Tl) crystal \cite{25Liu}.

\begin{figure*}
\includegraphics[width=0.8\textwidth]{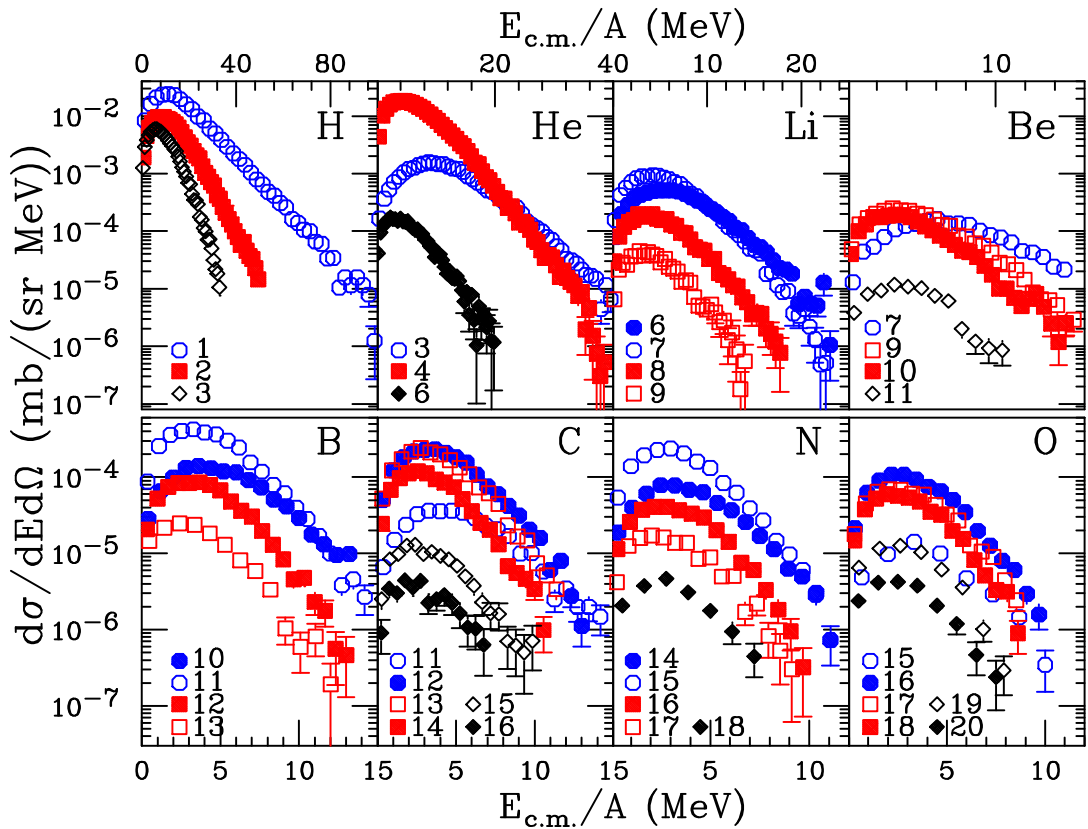}%fig2 goes here.
\caption{\label{fig:spectra}(Color online) Energy spectra of H, He, Li, Be, B, C, N, O from the central collisions of the reaction $^{112}$Sn+$^{112}$Sn. The mass number, A, of the isotopes is indicated in each panel. Isotopes with even A are shown in closed symbols, and isotopes with odd A are shown in open symbols. The energy spectra are obtained after background subtraction and efficiency correction.}
\end{figure*}

The center of mass energy spectra for isotopes from hydrogen to oxygen emitted in the $^{112}$Sn+$^{112}$Sn reactions are plotted in Fig. \ref{fig:spectra}. Except in the regions at high energies where statistics are poor, the energy spectra are very smooth. The loss of low energy particles due to lack of coverage at very forward angles is almost negligible. In general, the energy spectra drop exponentially with energy and the fragment cross sections drop with increasing charge. The high energy tails of the hydrogen isotopes are truncated due to the 6 cm length of the CsI(Tl) crystals of the LASSA device. The shapes of the corresponding energy spectra for the other Sn+Sn reactions are rather similar and are therefore not shown here. 

\begin{figure}
\includegraphics[width=0.5\textwidth]{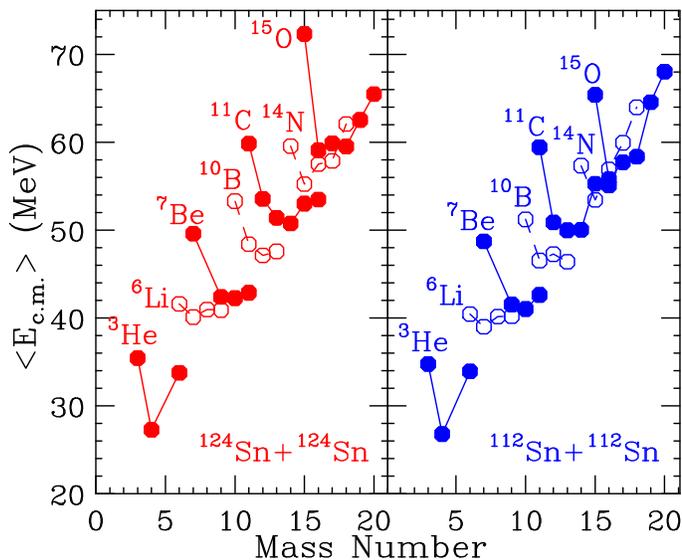}%fig3 goes here.
\caption{\label{fig:meanen}(Color online) Measured mean energies of isotopes as a function of mass number A.  Elements with even Z are shown with closed symbols, and elements with odd Z are shown with open symbols.}
\end{figure}

In Fig. \ref{fig:meanen} the measured mean energies of the isotopes are shown as a function of the mass number A. All the even Z (Z=2, 4, 6, 8) elements are represented by closed symbols and the odd Z (Z=1, 3, 5, 7) elements by open symbols. (The isotope mean energies for the $^{112}$Sn+$^{112}$Sn reaction, with the exception of $^{15}$O, have been published previously \cite{27Liu}.) There is an overall trend of increasing $\langle E_k \rangle$  with A. However, the proton-rich (N$<$Z) isotopes display significantly higher kinetic energies than the neutron-rich isotopes, meaning that that the energy spectra for these lighter isotopes have relatively smaller yields near the Coulomb barrier and larger yield at higher energies, leading to a higher apparent spectral temperature. The observation is consistent with the ``$^3$He-$^4$He puzzle'' \cite{28Gut, 29Renshaw, 30Neubert} where the mean kinetic energy of $^3$He is observed to be anomalously higher than that of $^4$He.

In our data, this ``anomalous energy puzzle'' extends beyond helium isotopes, up to oxygen. This trend is contrary to the behavior one would expect for emission from a single equilibrated source, where the mean kinetic energies of fragments are expected to increase with mass \cite{27Liu}. The higher kinetic energies of the lighter proton-rich isotopes has been quantitatively explained within the statistical Expanding Emitting Source model \cite{31Friedman} as a consequence of the low binding energies of the most proton-rich isotopes, such as $^3$He, $^7$Be, $^{10}$B, $^{11}$C and $^{15}$O, and the resulting large separation energies for these fragments from the emitting system \cite{27Liu, 32Xi}.  In these calculations, the large separation energies for these proton-rich fragments hinder their emission at later stages in the decay after the emitting system has expanded and cooled by expansion and emission \cite{27Liu, 32Xi}. The difference between the slopes of weakly bound proton-rich isotopes and their heavier counterparts has been used to extract the cooling curve or freeze-out temperature of hot multi-fragmenting systems \cite{32Xi, 33Wang}. Fig. \ref{fig:meanen} shows how this effect influences the total yields for 2$\leq$Z$\leq$8 fragments in $^{112}$Sn+$^{112}$Sn (right panel), and $^{124}$Sn+$^{124}$Sn (left panel) reactions. This appears to be a general trend, similarly exhibited by the other two reactions, $^{124}$Sn +$^{112}$Sn and $^{112}$Sn+$^{124}$Sn.

\section{\label{sec:t3he}t/$^3$He ratios}

The nuclear symmetry energy has contributions from both kinetic and potential energy. Both reduce the binding energy of systems with either neutron or proton excess. In the mean field limit, this implies that the poorly constrained symmetry mean field potential in a neutron-rich system should be repulsive to neutrons and attractive for protons. One expects, theoretically, that comparisons of neutron and proton observables such as the ratios of neutron and proton energy spectra should give direct information about the corresponding symmetry forces. However, measurements of neutron energies are difficult, requiring time of flight measurements with large, low efficiency scintillation arrays. In the limit of the coalescence approximation, the ratios of triton and $^3$He spectra would give information similar to that of the n/p measurements \cite{34Chen, 35Kohley}. Experimentally, it is much more preferable to detect charged particles such as triton and $^3$He. Even though the present experiment was not designed to optimize the detection of t/$^3$He yield ratios, the data nonetheless can be used to provide insights on future studies aimed to understand these ratios as they have been proposed as a probe to study the density dependence of symmetry energy \cite{2Li, 34Chen, 35Kohley}.

\begin{figure}
\includegraphics[width=0.5\textwidth]{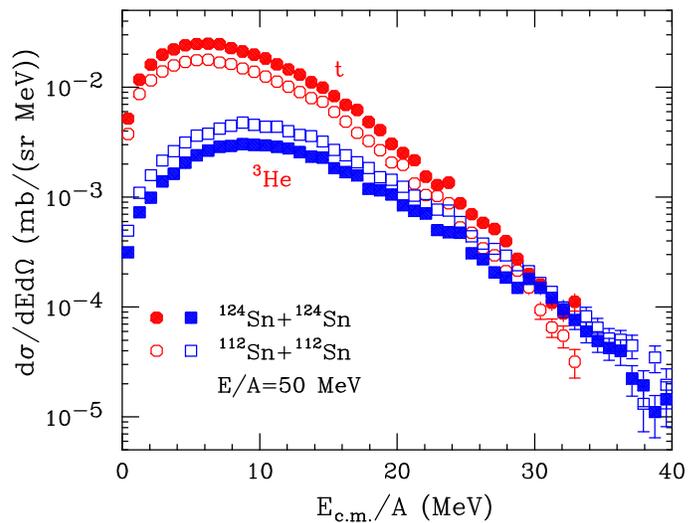}%fig4 goes here.
\caption{\label{fig:t3hespectra}(Color online) Energy spectra for central collisions.  Tritons are represented by circles and $^3$He are represented by squares.  Solid symbols are from the neutron-rich $^{124}$Sn+$^{124}$Sn reaction and open symbols are from the neutron-deficient $^{112}$Sn+$^{112}$Sn reaction.}
\end{figure}

 In Fig. \ref{fig:t3hespectra} the center of mass energy spectra are shown for tritons (open and closed circles) and $^3$He (open and closed squares) emitted in central collisions ($\hat{b}=b/b_{max}<0.2$). The solid symbols correspond to particles emitted in the neutron-rich $^{124}$Sn+$^{124}$Sn collisions while the open symbols correspond to particles emitted in the neutron-deficient $^{112}$Sn+$^{112}$Sn collisions. More triton particles, which are neutron-rich, are emitted from the $^{124}$Sn+$^{124}$Sn reaction as expected. Similarly, more $^3$He particles, which are neutron-deficient, are emitted from the $^{112}$Sn+$^{112}$Sn collisions. The shapes of the energy spectra for t and $^3$He are quite different especially near the Coulomb barrier. As the charge of $^3$He is twice that of $^3$H, the Coulomb barrier for $^3$He is higher than for $^3$H. This is clearly visible in Fig. \ref{fig:t3hespectra}, which shows that the maxima of the $^3$He energy spectra are shifted towards larger values.

For each isotope, the magnitudes of the differential multiplicities are reaction-dependent, but the shapes of the energy spectra from the $^{124}$Sn+$^{124}$Sn (closed symbols) and $^{112}$Sn+$^{112}$Sn (open symbols) reactions are rather similar. Such similarities are also observed in all the isotopes measured in this work as well as in the $^7$Li and $^7$Be spectra measured at E/A=35 MeV incident energy \cite{36Sun}. 

\begin{figure}
\includegraphics[width=0.5\textwidth]{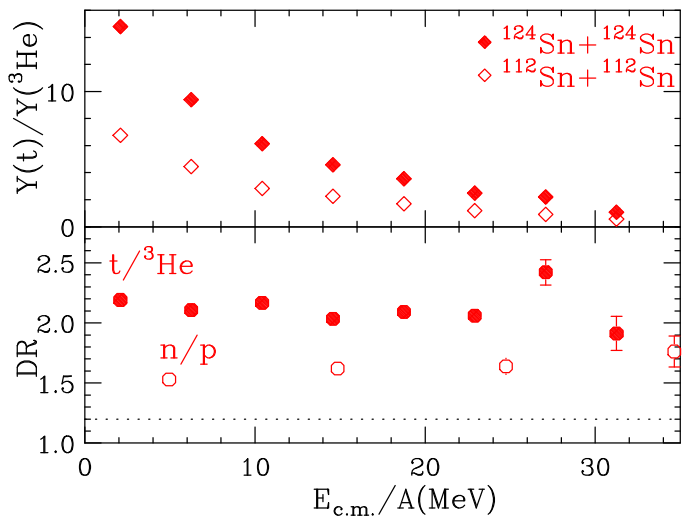}%fig5 goes here.
\caption{\label{fig:t3heYDR}(Color online) Top panel: Yield ratios of t and $^3$He (diamonds) plotted as a function of center of mass energy. Solid symbols are used for the neutron-rich $^{124}$Sn+$^{124}$Sn reaction and open symbols are used for the neutron-deficient $^{112}$Sn+$^{112}$Sn reaction.  Bottom panel:  Double yield ratios of t and $^3$He (closed circles) and n and p (open circles) plotted as a function of center of mass energy. The dotted line at 1.2 corresponds to the double ratio computed from the total number of neutrons divided by total number of protons in each reacting system.}
\end{figure}

Ratios of spectra can probe the isospin dependence with more sensitivity. The top panel of Fig. \ref{fig:t3heYDR} shows the yield ratios Y(t)/Y($^3$He) as a function of kinetic energy per nucleon for $^{124}$Sn+$^{124}$Sn (solid diamonds) and $^{112}$Sn+$^{112}$Sn (open diamonds). As expected, the ratios are larger for the neutron-rich system, $^{124}$Sn+$^{124}$Sn.  In both reactions, the Y(t)/Y($^3$He) ratios decrease rapidly with increasing kinetic energy. This is partly due to the different Coulomb barriers of the $^3$He and $^3$H affecting their energy spectra. 

Since Sn isotopes are used in both collision systems, the influence of the Coulomb force on the ratios of t/$^3$He spectra should be similar. However, we see a more rapid drop in the single yield ratios obtained from the neutron-rich system of $^{124}$Sn+$^{124}$Sn (solid diamonds) than those obtained from the $^{112}$Sn+$^{112}$Sn system (open diamonds). In a thermal picture, this indicates a larger value of the neutron chemical potential and a smaller value of the proton chemical potential in the neutron-rich $^{124}$Sn+$^{124}$Sn system than in the proton-rich $^{112}$Sn+$^{112}$Sn system. This is qualitatively consistent with effects expected from the symmetry mean field potential. 

Similar trends are observed in t/$^3$He ratios obtained in $^{48}$Ca+$^{48}$Ca and $^{40}$Ca+$^{40}$Ca systems at E/A = 80 MeV \cite{37Chajecki}. The large yield ratios at low kinetic energy are larger than the typical values from Boltzmann-Uehling-Uhlenbeck (BUU) models \cite{34Chen, 2Li} and values currently predicted by Improved Quantum Molecular Dynamical (ImQMD) model \cite{37Chajecki}. There is no definitive explanation why the large differences in the t/$^3$He yield ratios are not reproduced by transport model, but it may reflect the influence of cluster production. In a hybrid BUU-cluster production approach, Sobotka \textit{et al.} showed that the inclusion of alpha particles in the exit channel leads to an enhanced asymmetry of the lighter nucleons and clusters such as t and $^3$He \cite{38Sobotka}. In a similar vein, Natowitz \textit{et al.} have demonstrated that clusters, such as alpha particles, in the exit channel effectively increase the average symmetry energy in the final state, which will lead to an enhanced contribution of the symmetry potential to the effective chemical potentials that define the ratios of mirror nuclear yield such as t/$^3$He \cite{39Natowitz}. Clearly, the description of cluster production requires more attention to understand the role it plays in such effects. 

Following the methodology employed to study the neutron/proton yield ratios in ref. \cite{12Famiano}, a double ratio is constructed as DR(t/$^3$He)= (t/$^3$He)$_A$ /(t/$^3$He)$_B$, where A=$^{124}$Sn+$^{124}$Sn  and B=$^{112}$Sn+$^{112}$Sn, and is shown as the solid circles in the bottom panel of  Fig. \ref{fig:t3heYDR}. Over the regions with data, up to E/A=30 MeV, the double ratios are relatively flat, with values around 2.1. The experimental values are much larger than the no sensitivity limit at 1.2 dictated by conservation laws. 

For completeness, the double ratios DR(n/p) values from ref. \cite{12Famiano} are shown as open circles in the lower panel of Fig. \ref{fig:t3heYDR}.  Interestingly, the DR(t/$^3$He) values are larger than the DR(n/p) values at E$_{c.m.}$/A$<$30 MeV. This may reflect the large contributions of secondary decay to the neutron and proton spectra at low energies. Study of the n/p ratios using the ImQMD model suggests that to avoid clustering effects, one should use data at higher energies E$_{c.m.}>$25 MeV \cite{40Zhang}. Unfortunately the energy spectra for t and $^3$He clusters drop off exponentially and there are very few statistics beyond 30 MeV per nucleon as shown in the energy spectra. To obtain t and $^3$He data beyond E$_{c.m.}$/A=30 MeV, one would need to perform experiments for much longer time, with greater angular coverage, or at much higher incident energy. The differences in the DR(n/p) and DR(t/$^3$He) ratios suggest that the yield ratios of mirror nuclei such as t/$^3$He cannot be equated to the neutron/proton yield ratios without theoretical understandings of cluster formation and reaction dynamics, as well as their dependence on symmetry energy.

\begin{figure}
\includegraphics[width=0.5\textwidth]{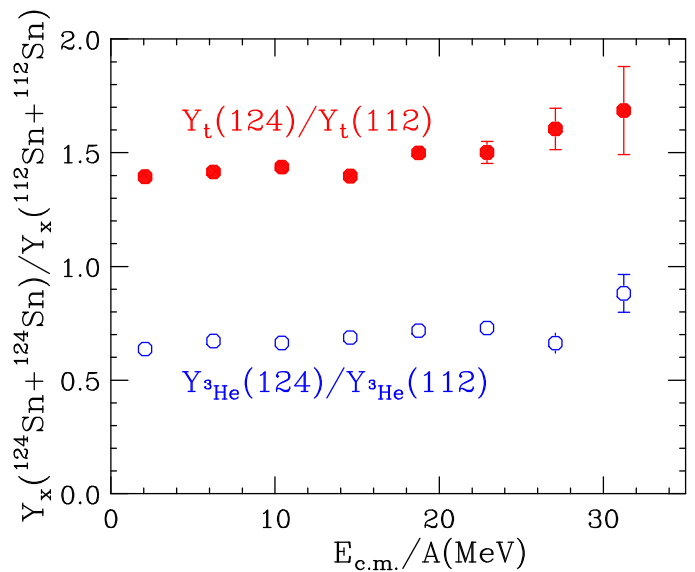}%fig6 goes here.
\caption{\label{fig:t3heY}(Color online) Yield ratios of tritons (solid circles) and $^3$He  (open circles).}
\end{figure}

To reduce the uncertainty due to the Coulomb and other pre-equilibrium effects and due to the possible efficiency problems in detecting different particles, we plot the yield ratios of an isotope from the two different reactions ($^{124}$Sn+$^{124}$Sn, $^{112}$Sn+$^{112}$Sn) instead of plotting the mirror ratios for single reactions. The triton yield ratios (solid circles) and $^3$He yield ratios (open circles) are shown in Fig. \ref{fig:t3heY}.  The average values of the triton ratios are 1.5; one expects that the neutron-rich tritons would be emitted in greater quantities from the neutron-rich $^{124}$Sn+$^{124}$Sn reaction system. The opposite is true for the neutron-deficient $^3$He isotopes, which have ratio values around 0.7. 

Both ratios seem to increase slightly with energy. Whether this continues to higher E$_{c.m.}$ cannot be answered with this data set, but could be more easily explored in experiments at higher incident energies. The corresponding ratios for the neutron-rich nuclei of (N-Z)=1 isotopes such as $^7$Li, $^{11}$B and $^{15}$N (not shown) are nearly the same at around 1.1, slightly lower than the values for triton ratios shown in Fig. \ref{fig:t3heY}. For the corresponding neutron-deficient mirror isotopes of (N-Z) = -1 such as $^7$Be, $^{11}$C and $^{15}$O (not shown) the measured ratios are around 0.6, similar to the $^3$He ratios. The observation of constant ratios is consistent with isoscaling \cite{14Geraci, 41Tsang}. The measured energy spectra of these heavy fragments do not extend much beyond E/A=12 MeV. Thus they do not shed light on the trend of the ratios as a function of energy or in the high energy region where the results can be better compared to the free neutron or proton observables.

\section{\label{sec:fragasym}fragment asymmetry}

\begin{figure}
\includegraphics[width=0.5\textwidth]{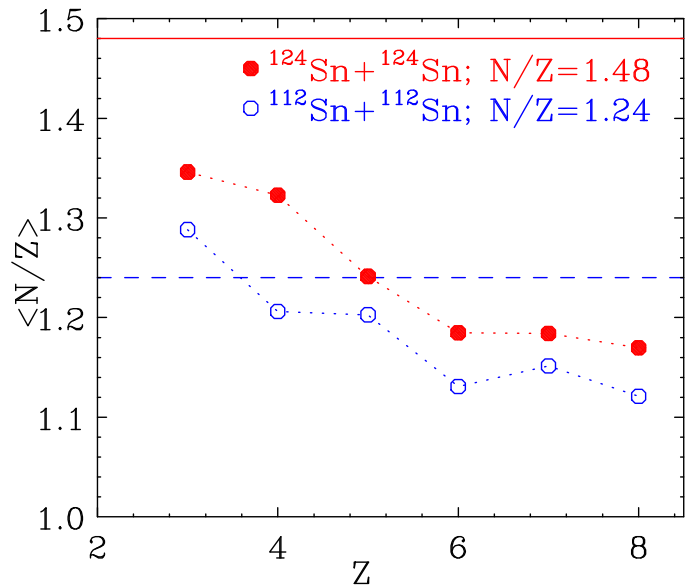}%fig7 goes here.
\caption{\label{fig:aveasym}(Color online) The average asymmetry $\langle N/Z \rangle$ as a function of Z.  The horizontal lines correspond to the initial $\langle N_0/Z_0 \rangle$ values.   The solid circles and solid line represent the $^{124}$Sn+$^{124}$Sn reaction and open circles and dashed line represent the $^{112}$Sn+$^{112}$Sn reaction. Dotted lines connecting the measured $\langle N/Z \rangle$ data are drawn to guide the eye.}
\end{figure}

More recently the average asymmetry, $\langle N/Z \rangle$, values of fragments of a given Z have been proposed as observables that are complementary to the free nucleon yield ratios of Y(n)/Y(p) in providing information about symmetry energy \cite{42Colonna}. This quantity is calculated as a weighted sum over yields measured in the angular domain $70^{\circ}< \theta_{c.m.}<110^{\circ}$ as follows:
\begin{equation}
\langle N/Z \rangle=\bigg[\frac{\sum_i N_i \times Y(Z,N_i )}{\sum_i Y(Z,N_i ) }\bigg] /Z 
\label{eq:aveasym}
\end{equation}
Here, $Y(Z,N_i)$ is the yield of fragments of charge Z and neutron number $N_i$ observed experimentally in this angular domain. Some of the results shown in Fig. \ref{fig:aveasym} have been published in \cite{16Das, 15Liu}. We include a more thorough discussion on this observable for completeness in view of the next new asymmetry observable, $\Sigma$N/$\Sigma$Z, and to reexamine its virtues and limitations in placing constraints on the symmetry energy.  

The average asymmetry $\langle N/Z \rangle$ is shown in Fig. \ref{fig:aveasym} as a function of the fragment charge number Z. As expected, the  $\langle N/Z \rangle$ values are larger for the more neutron-rich system (solid circles). However, the observed  $\langle N/Z \rangle$ values of the fragments for the neutron-rich system are much lower than the initial N/Z value of the projectile and target,  $\langle N_0/Z_0 \rangle$ =1.48 (solid horizontal line near the top of the figure). For the neutron-deficient system, the fragment  $\langle N/Z \rangle$ values (open circles) lie much closer to the $\langle N_0/Z_0 \rangle$ value of 1.24 (dashed horizontal line).  The observed trends can be explained in statistical fragmentation models that include sequential decays \cite{16Das}. 

Since the energy spectra are exponential in shape, the $\langle N/Z \rangle$ values mostly reflect the N/Z values of the low-energy fragments. The general trend of the N/Z values is similar for the two reactions except for Z=4.  Because $^8$Be is unstable and decays into two alpha particles, it is not detected in the experiment. The effect of $^8$Be will be discussed in more detail later in this section. 

In a recent study using the Stochastic Mean Field (SMF) model, Colonna et al. simulated the Sn reactions studied here and suggested the construction of the observable $\Sigma$N/$\Sigma$Z as a function of the fragment kinetic energies as an observable to study isospin effects and the symmetry energy in heavy ion collisions \cite{42Colonna}. Technically, $\Sigma$N/$\Sigma$Z is computed as follows:
\begin{equation}
\sum N \big/\sum Z = \frac{\sum_{i,j} N_i \times Y(Z_j,N_i )}{ [ \sum_{i,j} Z_j \times Y(Z_j,N_i ) ] }
\label{eq:sNsZ}
\end{equation}

Here, $Y(Z_j,N_i)$ is the yield of fragments of charge $Z_j$ and neutron number $N_i$ observed in this angular domain experimentally.  Note that this is equivalent to Eq. \ref{eq:aveasym} in the case when only one element is included in the sums.

\begin{figure}
\includegraphics[width=0.5\textwidth]{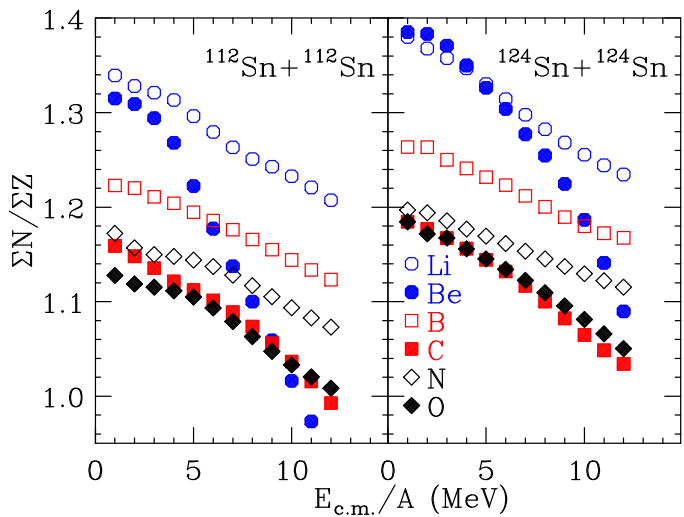}%fig8 goes here.
\caption{\label{fig:sNsZ}(Color online) Measured $\Sigma$N/$\Sigma$Z of fragments as a function of center of mass kinetic energy per nucleon. The $^{112}$Sn+$^{112}$Sn reaction is shown in the left panel and the $^{124}$Sn+$^{124}$Sn reaction is shown in the right panel. Open symbols are used for elements with odd Z and closed symbols are used for elements with even Z.}
\end{figure}

The effects of unstable nuclei produced as the primary fragments can be illustrated using the case of Be. In Fig. \ref{fig:sNsZ}, the $\Sigma$N/$\Sigma$Z values of the individual elements are plotted as a function of the kinetic energy per nucleon in the center of mass frame. The $\Sigma$N/$\Sigma$Z values for all elements decrease with kinetic energy. The trends exhibited by the elements are similar except for Be. This reflects the fact that $^8$Be decays were not included in the measured observable. (In principle, one can identify $^8$Be but this was not done because the current experiment was not designed to detect $^8$Be.) Other elements have particle unstable ground and excited state nuclei, but Be is unique in that the N=Z isotope $^8$Be is strongly produced and strongly fed by secondary decay, but is unstable. Thus, the energy dependence of $\Sigma$N/$\Sigma$Z drops much more rapidly for Z=4 fragments than for the other elements. The same phenomenon is observed both in $^{112}$Sn+$^{112}$Sn (left panel) and $^{124}$Sn+$^{124}$Sn (right panel) systems. However, the drop is steeper for the $^{112}$Sn+$^{112}$Sn reaction. 

\begin{figure}
\includegraphics[width=0.5\textwidth]{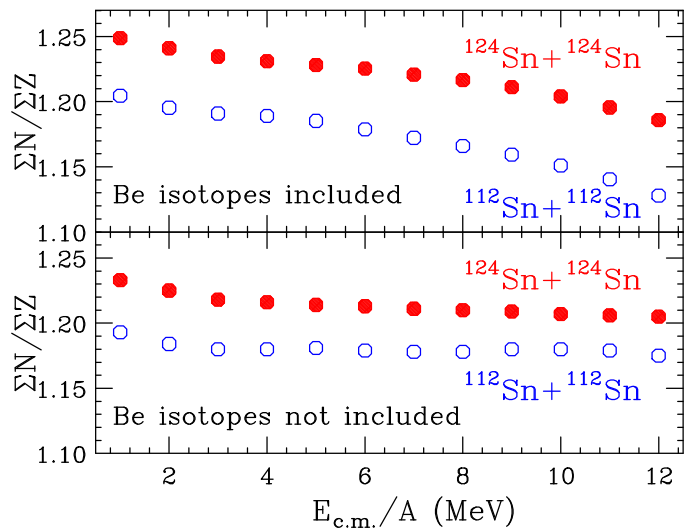}%fig9 goes here.
\caption{\label{fig:benobe}(Color online) Measured $\Sigma$N/$\Sigma$Z of fragments, as a function of center of mass kinetic energy per nucleon. Closed circles represent the $^{124}$Sn+$^{124}$Sn reaction and open circles represent the $^{112}$Sn+$^{112}$Sn reaction.  The measured $\Sigma$N/$\Sigma$Z including Be isotopes are shown in the top panel; measured $\Sigma$N/$\Sigma$Z omitting Be isotopes are shown in the bottom panel.}
\end{figure}

Fig. \ref{fig:benobe} shows $\Sigma$N/$\Sigma$Z as a function of the kinetic energy per nucleon for the $^{124}$Sn+$^{124}$Sn (closed circles) and $^{112}$Sn+$^{112}$Sn (open circles) systems. Here, all of the observed isotopes with $3\leq Z_j \leq 8$ have been included. As expected, the fragment $\Sigma$N/$\Sigma$Z values are higher for the neutron-rich $^{124}$Sn+$^{124}$Sn system, but the overall effect is rather small at about 4\% higher than the neutron-deficient  $^{112}$Sn+$^{112}$Sn system. Furthermore, the trends of the ratios are very sensitive to whether Be isotopes are included in the construction of $\Sigma$N/$\Sigma$Z, especially at high center of mass energies. When experimentally-measured Be isotopes are included as shown in the top panel, both ratios decrease with energy. However, when Be isotopes are excluded as shown in the bottom panel, $\Sigma$N/$\Sigma$Z ratios are nearly flat as a function of energy. 

\begin{figure}
\includegraphics[width=0.5\textwidth]{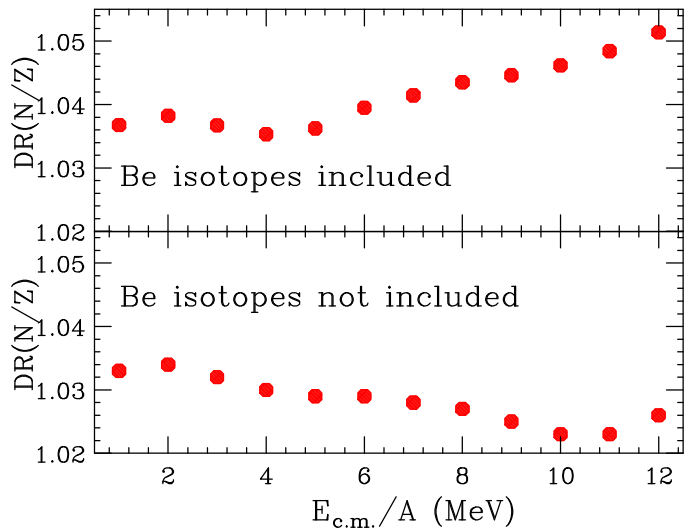}%fig10 goes here.
\caption{\label{fig:DRbenobe}(Color online) Double ratios DR(N/Z) as a function of kinetic energy.  The measured ratios including Be isotopes are shown in the top panel; measured ratios omitting Be isotopes are shown in the bottom panel.}
\end{figure}

Double ratios have been used in comparisons of neutron and proton spectra to remove the influence of effects such as inefficiencies in neutron detection \cite{12Famiano}, so it is natural to explore whether double ratios could correct for the inefficiency in $^8$Be detection. Unfortunately, it appears that the double ratios, DR(N/Z)=($\Sigma$N/$\Sigma$Z)$_A$/($\Sigma$N/$\Sigma$Z)$_B$, where A=$^{124}$Sn+$^{124}$Sn  and B=$^{112}$Sn+$^{112}$Sn, are rather sensitive to the inclusion of the Be isotopes.  The top panel of Fig. \ref{fig:DRbenobe} shows that the experimental double ratios increase with the kinetic energy beyond E/A$>$4 MeV when detected Be isotopes are included in the analysis. This increasing trend changes to a slightly decreasing trend when the Be isotopes are excluded (bottom panel). Thus, one needs to accurately model the secondary decay of the isotopically-resolved fragments quantitatively to reproduce the isospin effects of either the single ratios $\Sigma$N/$\Sigma$Z or the double ratios DR(N/Z)=($\Sigma$N/$\Sigma$Z)$_A$/($\Sigma$N/$\Sigma$Z)$_B$ at the level of a few percent in order to constrain the density dependence of the symmetry energy using this observable. At present, DR(N/Z) predicted by transport calculations depends on the secondary decay models \cite{42Colonna} and will need much more developmental work on sequential decays before this observable can be used to extract information about symmetry energy.

\section{\label{sec:summary}summary}

In summary, we have measured the energy spectra of emitted charged particles from Z=1 to Z=8 in central collisions of Sn isotopes at E/A=50 MeV incident energy. The average kinetic energies of the proton-rich as well as N=Z isotopes such as $^3$He, $^7$Be, $^{10}$B, $^{11}$C, $^{14}$N, and $^{15}$O are higher than those of their corresponding neutron-rich isotopes, an effect often described as the ``$^3$He puzzle''. We note that this trend has been quantitatively reproduced by the statistical Expanding Emitting Source model, where it was shown to stem from the suppression of the emission of weakly bound proton-rich isotopes during the later stages of the decay after the system has expanded and cooled. We construct yield ratios such as  Y(t)/Y($^3$He) as a function of kinetic energy. These ratios are not the same as the Y(n)/Y(p) at low kinetic energy, E/A$<$25 MeV. To minimize the contributions from Coulomb effects, sequential decays, and cluster formation, measurements of the fragment observables with higher statistical accuracies from reactions at higher incident are desirable. The results at high kinetic energies are needed to determine if Y(t)/Y($^3$He) can indeed be used to substitute Y(n)/Y(p) in the study of the sensitivity to symmetry energy. We also explore average asymetry, $\langle N/Z \rangle$, ratios constructed from intermediate mass fragments. Sequential decays tend to push the $\langle N/Z \rangle$ values from very asymmetric systems closer to each other and further from the initial N/Z values of the composite system formed by the projectile and target.  The small measured values coupled with the importance of sequential decays in $\langle N/Z \rangle$ as well as $\Sigma$N/$\Sigma$Z suggest that more theoretical study is needed to understand these observables and accurate sequential decay models be developed before they can be employed to constrain the density dependence of symmetry energy. 

\section*{\label{sec:ack}Acknowledgement}

This work has been supported by the U.S. National Science Foundation under Grants PHY 060007 (MSU) and the Department of Energy under grant numbers DE-FG02-87ER-40316 (WU) and DE-FG02-88ER-40404 (IU).

% The \nocite command causes all entries in a bibliography to be printed out
% whether or not they are actually referenced in the text. This is appropriate
% for the sample file to show the different styles of references, but authors
% most likely will not want to use it.
%\nocite{*}

\bibliography{txlu_prc}% Produces the bibliography via BibTeX.

\end{document}